\documentclass[prl,aps,reprint,superscriptaddress,longbibliography]{revtex4-1}
\usepackage{graphics}
\usepackage{epsfig}
\usepackage{amsmath}
\usepackage{amsfonts}
\usepackage{amssymb}
\usepackage{graphicx}
\usepackage{color}
\usepackage{tabularx}
\usepackage{txfonts}

\begin{document}

\title{Doping induced site-selective Mott insulating phase in LaFeO$_3$}
\author{S. Jana}
\email{somnath.jana@physics.uu.se}
\altaffiliation{Present Address: Institute for Methods and Instrumentation in Synchrotron Radiation Research FG-ISRR, Helmholtz-Zentrum Berlin f{\"u}r Materialien und Energie, Albert-Einstein-Strasse 15, 12489 Berlin, Germany}
\affiliation{Department of Physics and Astronomy, Uppsala University, 752 36 Uppsala, Sweden}
\affiliation{Solid State and Structural Chemistry Unit, Indian Institute of Science, Bengaluru 560012, India}
\author{S. K. Panda}
\affiliation{Centre de Physique Th{\'e}orique, Ecole Polytechnique, CNRS UMR 7644, Universit{\'e} Paris-Saclay, 91128 Palaiseau, France}
\author{D. Phuyal}
\affiliation{Department of Physics and Astronomy, Uppsala University, 752 36 Uppsala, Sweden}
\author{B. Pal}
\affiliation{Solid State and Structural Chemistry Unit, Indian Institute of Science, Bengaluru 560012, India}
\author{S. Mukherjee}
\altaffiliation{Present Address: Department of Physics and Astronomy, Uppsala University, 752 36 Uppsala, Sweden}
\affiliation{Solid State and Structural Chemistry Unit, Indian Institute of Science, Bengaluru 560012, India}
\author{A. Dutta}
\affiliation{Solid State and Structural Chemistry Unit, Indian Institute of Science, Bengaluru 560012, India}
\author{P. Anil Kumar}
\altaffiliation{Present Address: Seagate Technology, 1 Disc Drive, Springtown, Northern Ireland BT48 0BF, United Kingdom}
\affiliation{Department of Engineering Sciences, Uppsala University, 752 36 Uppsala, Sweden}
\author{D. Hedlund}
\affiliation{Department of Engineering Sciences, Uppsala University, 752 36 Uppsala, Sweden}
\author{J. Sch\"ott}
\affiliation{Department of Physics and Astronomy, Uppsala University, 752 36 Uppsala, Sweden}
\author{P. Thunstr\"om}
\affiliation{Department of Physics and Astronomy, Uppsala University, 752 36 Uppsala, Sweden}
\author{Y. Kvashnin}
\affiliation{Department of Physics and Astronomy, Uppsala University, 752 36 Uppsala, Sweden}
\author{H. Rensmo}
\affiliation{Department of Physics and Astronomy, Uppsala University, 752 36 Uppsala, Sweden}
\author{M. Venkata Kamalakar}
\affiliation{Department of Physics and Astronomy, Uppsala University, 752 36 Uppsala, Sweden}
\author{Carlo. U. Segre}
\affiliation{CSRRI \& Department of Physics, Illinois Institute of Technology, Chicago, IL 60616, USA}
\author{P. Svedlindh}
\affiliation{Department of Engineering Sciences, Uppsala University, 752 36 Uppsala, Sweden}
\author{K. Gunnarsson}
\affiliation{Department of Engineering Sciences, Uppsala University, 752 36 Uppsala, Sweden}
\author{S. Biermann}
\affiliation{Centre de Physique Th{\'e}orique, Ecole Polytechnique, CNRS UMR 7644, Universit{\'e} Paris-Saclay, 91128 Palaiseau, France}
\affiliation{Coll{\`e}ge de France, 11 place Marcelin Berthelot, 75005 Paris, France}
\author{O. Eriksson}
\affiliation{Department of Physics and Astronomy, Uppsala University, 752 36 Uppsala, Sweden}
\affiliation{School of Science and technology, {\"O}rebro University, SE-70182 {\"O}rebro, Sweden}
\author{O. Karis}
\affiliation{Department of Physics and Astronomy, Uppsala University, 752 36 Uppsala, Sweden}
\author{D. D. Sarma}
\email{sarma@iisc.ac.in}
\affiliation{Solid State and Structural Chemistry Unit, Indian Institute of Science, Bengaluru 560012, India}
%\begin{abstract}
%\end{abstract}

\maketitle

{\bf Tailoring transport properties of strongly correlated electron systems in a controlled fashion counts among the dreams of materials scientists. In copper oxides, varying the carrier concentration is a tool to obtain high-temperature superconducting phases. In manganites, doping results in exotic physics such as insulator-metal transitions (IMT), colossal magnetoresistance (CMR), orbital- or charge-ordered (CO) or charge-disproportionate (CD) states. In most oxides, antiferromagnetic order and charge-disproportionation are asssociated with insulating behavior. Here we report the realization of a unique physical state that can be induced by Mo doping in LaFeO$_3$: the resulting metallic state is a site-selective Mott insulator where itinerant electrons evolving in low-energy Mo states coexist with localized carriers on the Fe sites. In addition, a local breathing-type lattice distortion induces charge disproportionation on the latter, without destroying the antiferromagnetic order. A state, combining antiferromangetism, metallicity and CD phenomena is rather rare in oxides and may be of utmost significance for future antiferromagnetic memory devices.} 

Ever since the pioneering work by Mott~\cite{0370-1298-62-7-303}, IMTs in strongly correlated systems remain an area of extensive investigations. The commonly employed strategies to realize IMT~\cite{RevModPhys.70.1039} are by controlling the bandwidth of the $d$ bands using pressure or by the introduction of charge carriers using chemical substitution into either Mott-Hubbard or charge-transfer insulators (CTI) in the Zaanen-Sawatzky-Allen phase diagram~\cite{PhysRevLett.55.418}. Doping remains a reliable means to induce IMTs, and has attracted much attention, following the demonstration of high temperature cuprate superconductors~\cite{RevModPhys.78.17}.
\par
Doping perovskite transition metal oxides $ABO_3$ results in exotic physics, such as IMT, CMR, orbital and charge ordered states. The $A$-site ions, usually a rare-earth or alkali-metal, provide the structural skeleton after donating electrons to the valence band, while $B$-site ions, usually a transition metal, and the oxygen ions are predominantly responsible for the low-energy electronic states. Often the $A$-site is substituted to either introduce holes/electrons or to induce structural changes for manipulating material properties. In particular in manganites, a large number of materials with chemical formula $R_{1-x}A_x$MnO$_3$ ($R$ = La, Pr, or Nd and $A$ = Ca, Br, Sr, Pb)~\cite{cmr_review,Coey_review} have been realized, where most of them exhibit CMR for $x$$\sim$0.3~\cite{RevModPhys.73.583}, around the IMT where doping transforms an antiferromagnetic (AFM) charge transfer insulator to a ferromagnetic (FM) mixed-valence metallic compound~\cite{PhysRevB.51.14103}. The physics of this transition has been attributed to the double-exchange mechanism~\cite{PhysRev.82.403,PhysRev.100.675} which promotes FM ordering by allowing the $e_g$ electron to hop between Mn$^{3+}$ and Mn$^{4+}$ ions, resulting in metallic conduction, while the insulating AFM state in the parent compound arises due to the super-exchange interactions between the Mn ions. At larger values of $x$ ($>$ 0.5), the metallic FM ground state transforms into a CO ground state where the Mn$^{3+}$ and Mn$^{4+}$ ions order spatially in the crystalline lattice~\cite{PhysRevLett.76.3188}, suppressing the double-exchange and thus favouring AFM ordering due to super-exchange interaction~\cite{PhysRevLett.78.543}. Thus the insulating state always coexists with the AFM ordering and metallic conduction appears to be a prerequisite for the ferromagnetism in these compounds. 

In strong similarity with manganites, the perovskite ferrites $A$FeO$_3$ ($A$ = Lu, Eu, Y, Pr, and La)~\cite{PhysRevLett.92.037202,PhysRevB.62.844} are typically high-spin (Fe$^{3+}$, $S$ = 5/2), wide-gap insulators with AFM ground states~\cite{RFeO3_White,PhysRevB.64.094411}. Several attempts have been made to realise IMT and other phenomena in ferrites too, following the route of $A$-site substitution~\cite{PhysRevLett.79.297, PhysRevB.48.14818}. Among the most studied ferrites, LaFeO$_3$ has been hole-doped, but unlike the corresponding manganite, La$_{1-x}$Sr$_x$FeO$_3$ retains an insulating AFM state even up to 40\% Sr substitution, establishing the robustness of the antiferromagnetism in this compound~\cite{PhysRevB.48.14818}. 
At higher doping around $x$ = 0.7, CD occurs on the Fe sites (Fe$^{3+}$ and Fe$^{5+}$) at low temperature, resulting in an insulating AFM ground state~\cite{PhysRevLett.98.126402}. $B$-site doping has also been explored~\cite{PhysRevLett.80.4004,PhysRevB.49.14238,0022-3727-44-10-105401,0953-8984-8-43-001}. Idrees \emph{et al.}~\cite{0022-3727-44-10-105401} report that even 50\% Ni doping in LaFeO$_3$ (LaFe$_{1-x}$Ni$_x$O$_3$) could not destabilize an insulating AFM state. 

\begin{figure*}[t]
\includegraphics[width=0.99\textwidth]{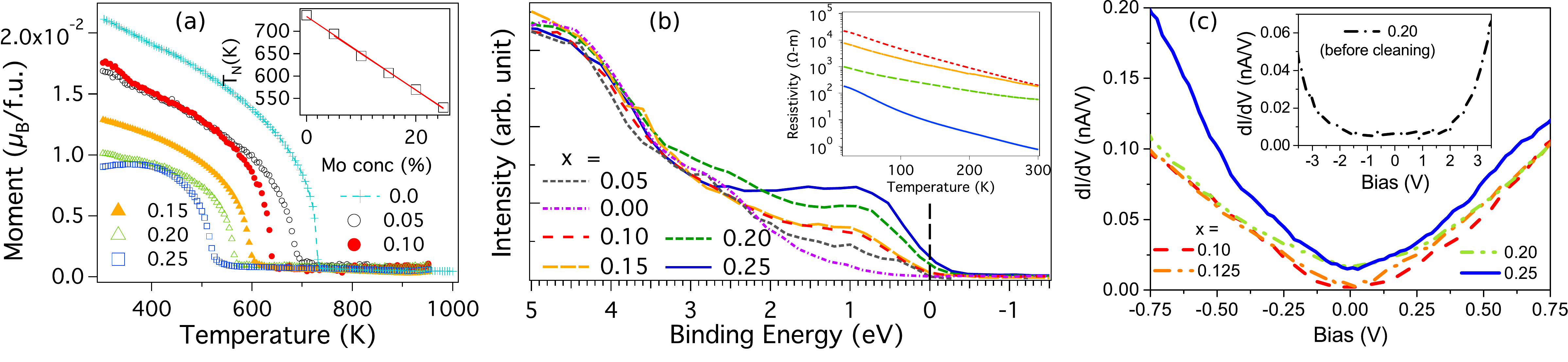}
\caption{{\bf Magnetic and electronics properties of doped LaFeO$_3$.} (a) Temperature dependence of the magnetization for all investigated compositions. Inset: The variation of antiferromagnetic transition temperature (T$_{N}$) as a function of doping ($x$). (b) Valence band photo-electron spectra, normalized to the La $5p_{3/2}$ peak ($\sim$17.7 eV) for all the compositions. Inset: The resistivity vs temperature plots for $x$ = 0.10, 0.15, 0.20, 0.25. (c) Scanning tunnelling spectroscopic (STS) data collected at room temperature in an UHV system using a variable temperature STM system. Standard ac modulation technique was used for STS measurements with an ac modulation amplitude of 10 mV and frequency 2731 Hz. The dI/dV spectra for $x$ = 0.10, 0.125, 0.20, 0.25 compositions are plotted. Inset shows the same for uncleaned $x$ = 0.20 sample. Tunnel spectra were taken at different locations over 2x2 mm$^2$ area on each composition and the plotted spectra is the spatial average of all of them.}
\label{Fig1}
\end{figure*}
The coexistence of AFM order and metallicity is extremely rare in perovskite oxides~\cite{Guo2018}. However the search for such an exotic state has continued to draw significant attention since the discovery of high-$T_{c}$ superconductivity in copper oxides. Antiferromagnets are of interest in spintronic applications~\cite{RevModPhys.90.015005,Jungwirth2018,Wadley587} due to their fast dynamics, stability in external magnetic field, where itinerant electrons allow for  manipulating and reading the magnetic states~\cite{Higo2018,Lebrun2018}. Since a few studies indicate the robustness of  antiferromagnetism in LaFeO$_3$~\cite{PhysRevB.48.14818,0022-3727-44-10-105401}, the possibility of a metallic AFM state could be explored through appropriate $B$-site substitution. Since $4d$ elements such as Mo have much broader $d$ bands, this comes as a natural choice.

To obtain an IMT, without an associated magnetic transition in a wide-gap perovskite oxide, this work focuses on electron doping in bulk LaFe$_{1-x}$Mo$_x$O$_3$ (LFMO-$x$). Based on magnetic, transport, X-ray spectroscopy measurements and electrnic structure calculations, we investigate the variation of electronic and magnetic properties as a function of $x$ and establish that the system shows site-selective Mott insulating state for $x$ $\ge$ 0.20. The theoretical analysis of the 25\% Mo doped system further proposes CD in the Fe-sites, where two different types of Fe ions, assuming 2+ and 3+ valence states, form the insulating network, while the partially filled Mo states form an isolated band that provides metallic character at the Fermi level. The proposed CD state has also been verified from X-ray absorption spectroscopy (XAS) and X-ray absorption fine structure analysis (XAFS), providing conclusive evidence for a unique ground state where AFM spin-order, site-selective Mott insulating behavior and CD phenomena coexist.

\begin{figure*}[t]
\includegraphics[width=1.99\columnwidth]{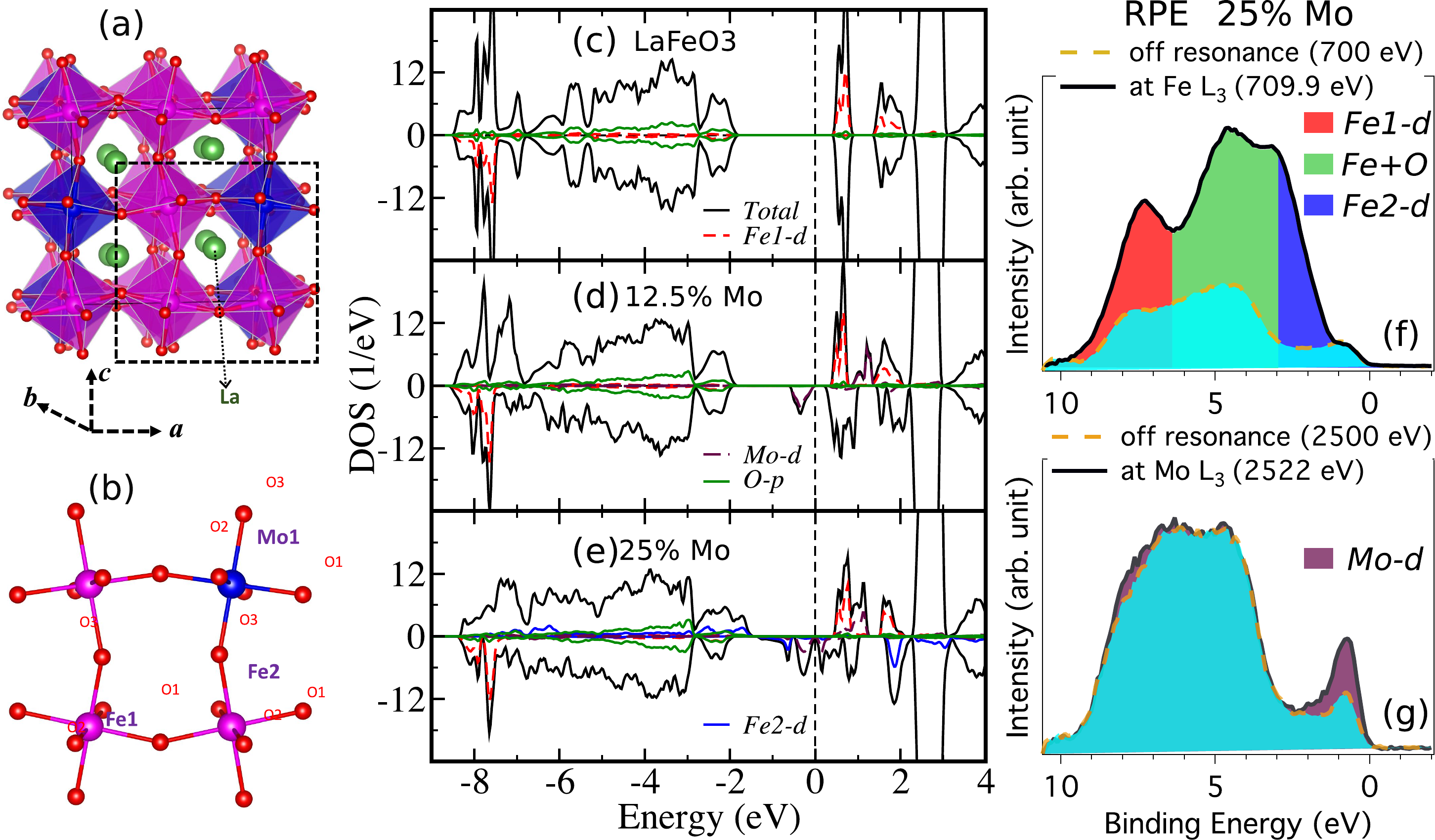}
\caption{{\bf Theoretically obtained relaxed crystal geometry, density of states and experimental resonant photo emission spectra (RPES).} (a) The schematic of the $\sqrt(2) \times 1 \sqrt(2)$ super-cell, considered for theoretical simulations of $x$ = 0.25 sample, consisting 6 Fe and 2 Mo ions. The Mo ions are substitutes at the Fe sites and this particular arrangement of Mo ions is energetically favorable (see SI). (b) The dotted part of the structure in panel (a) is highlighted for a better visualization of the two kinds of Fe-ions (Fe1, and Fe2), present in the relaxed structure of 25$\%$ doped sample. Our optimized structure within DFT+U approach find that all the six bonds in Fe1-O$_6$ octahedra are almost equal to their corresponding values in pure LaFeO$_3$ (2.02 \AA), while fours Fe-O bonds are larger (2.07 \AA) in Fe2O$_6$ octahedra, indicating a breathing-type lattice distortion. DFT+U total (black solid line), partial density of states of the Fe1-$d$ (red solid line), Mo-$d$ (maroon dotted line), O-$p$ (blue solid line), and Fe2-$d$ (blue solid line) in the AFM ground state of LaFe$_{1-x}$Mo$_x$O$_3$ corresponding to (c) $x$ = 0.0, (d) $x$ =0.125, and (e) $x$ = 0.25. The Fermi level for LaFeO$_3$ has been assigned such a way that Fe-$d$ states appear at a binding energy, consistent with our experimental resonance photoemission spectra (RPES). The RPES spectra obtained at (f) Fe-$L_3$ and (g) Mo-$L_3$ edge for $x$ = 0.25 composition. The off and on resonance spectra for Fe (Mo) were collected when the excitation energy was set at 700 eV (2500 eV) and 709.9 eV (2522 eV), respectively. The important spectral features of the electronic states of panel (e) are identified from RPES data.}
\label{VB_Structure_DOS_RPES}
\end{figure*}

\section*{Results}
Six samples of LFMO-$x$ ($x$ = 0.00, 0.05, 0.10, 0.15, 0.20, 0.25) are studied. 
Phase purity and the highly crystalline nature of the samples were confirmed by X-ray diffraction (see Supplementary Information (SI)).
Fig.\ \ref{Fig1}(a) shows the magnetization (M) as a function of temperature for all compositions. The data is measured while cooling in an applied field of 1000 Oe. The AFM transition temperature ($T_{N}$) of LaFeO$_3$ agrees well with the previously reported value \cite{LaFeO3_FC}. The $T_{N}$ is observed to decrease linearly with increasing Mo doping (see inset of Fig.\ \ref{Fig1} (a)), while the ferromagnetic (FM) like saturation of the moment below $T_{N}$ is typically observed due to the small canting of the moments at relatively high applied field \cite{LaFeO3_FC}. These results indicate that even at the highest doping, our system remains antiferromagnetic. 

Next, the spectral function at $E_F$ is examined by means of valence band photoelectron spectroscopy (PES) conducted at room temperature (RT) as displayed in Fig.\ \ref{Fig1}(b). All spectra are energy calibrated relative to a Au reference measured under identical conditions and intensities are normalized relative to the La $5p_{3/2}$ peak (17.7 eV). For the undoped sample, a large gap between $E_F$ and the top of the valence band is present, while with the Mo doping, electronic states start to appear in the gap. For $x$ = 0.05, 0.10 and 0.15 samples, the tail of the valence band is ending below $E_F$, i.e.\ no DOS at $E_F$. However, for $x$ = 0.20 and 0.25 samples, clearly, a sizeable number of states appear across $E_F$, establishing the AFM metallic state at room temperature for $x\geq$ 20\%. 

The resistivity ($\rho$) data as displayed in the inset of Fig.\ \ref{Fig1}(b) reveals a two order of magnitude reduction in the magnitude for LFMO-0.25 compared to the LFMO-0.10 composition. We note that resistivity of the pure LaFeO$_3$ is as high that we could not measure it due to the limitation in our set-up.  The major takeaway of our measurement is that despite negative value of $\partial\rho$/$dT$, the $\rho$(T) of LFMO-0.25 at low T does not diverge and saturates to a rather small value. The negative $\partial\rho$/$dT$ is not very surprising as also has been reported in other metallic systems like ferromagnetic CrO$_2$~\cite{aplCrO2} and antiferromagnetic CaCrO$_3$~\cite{KomarekCaCrO3,PhysRevB.83.165132} which is considered to be in the crossover regime between localized and itinerant electrons~\cite{PhysRevB.78.054425} alike our system. The reason has been attributed to the domination of the insulating grain boundaries in the polycrystalline sample~\cite{KomarekCaCrO3,aplCrO2}. We will illustrate below that this is indeed the case for our system.

In order to provide further credence to the observed IMT from the XPS data, the scanning tunnelling spectroscopy (STS) measurements is displayed in Fig.\ \ref{Fig1}(c). This shows a finite $dI/dV$ at zero bias $(V = 0)$ for both 20\% and 25\% Mo doping, indicates finite DOS at the Fermi level ($E_F$) for both samples, while negligible $dI/dV$ at $V = 0$ for 10\% and 12.5\% Mo doping, providing another proof in favor of the IMT at $x$ = 0.20. Inset of Fig.\ \ref{Fig1}(c), showing the $dI/dV$-$V$ curve measured on the uncleaned 20\% Mo doped sample, demonstrates the insulating nature of the grain boundary, as dI/dV remains negligibly small for large range of V. Complete suppression of the metallic character at the grain boundary is an indication of high sensitivity to the defects. This further explain the observed nature of our $\rho$(T) data.  

To understand the mechanism of the IMT, we calculate the electronic structure of $x$ = 0.00, 0.125 and 0.25 samples by means of DFT+U (density functional theory + Hubbard $U$) approach as implemented in the Wien2k code~\cite{wien2k,FPLAPW_Wien2k}. The technical details of these calculations are provided in the SI. Experimentally, LaFeO$_3$ ($x$ = 0.00) forms in the orthorhombic crystal structure with space group $Pnma$~\cite{TAGUCHI2005773}. We consider a $\sqrt{2}$$\times$1$\times$$\sqrt{2}$ cell to simulate two different doping concentrations. The cell for $x$ = 0.25 (6 Fe and 2 Mo) is displayed in Fig.~\ref{VB_Structure_DOS_RPES}(a). This specific arrangement of the Mo ions are found to be lowest in energy (see SI), while Mo replaces Fe completely disorderedly as probed by XRD and EXAFS (see SI). The optimized structure reveals that Mo doping produces two types of Fe atoms (Fe1 and Fe2 in Fig.~\ref{VB_Structure_DOS_RPES}(b)). The Fe1 is surrounded by only Fe ions as in pure LaFeO$_3$ and thus all the six Fe1-O bond distances remain almost equal to their corresponding values in pure compound (2.02~\AA), while in the Fe2O$_6$ octahedra, the presence of nearest Mo ions makes Fe2-O2 and Fe2-O3 (total 4 bonds) distances equal to 2.07~\AA. Such breathing-type lattice distortion has been attributed as the cause of CD state in many rare-earth nickelate perovskites~\cite{PhysRevLett.82.3871,PhysRevB.78.212101} and also in CaFeO$_3$~\cite{PhysRevB.62.844}. Since a very similar Fe-O bond disproportionation or {\em breathing distortion} is observed in the present system, it strongly indicates the possibility of such novel state.

We analyze the total and projected DOS  of the AFM ground state for both the pure and two doped systems. We note that the AFM-I state is found to be the lowest energy state as described in the SI. The total DOS of pure LaFeO$_3$, displayed in Fig.~\ref{VB_Structure_DOS_RPES}(c), reveals that the magnetic ground state is a wide gap insulator. The calculated band gap is 2.25 eV, which is in good agreement with the experimental value of 2.0-2.2 eV~\cite{PhysRevB.48.14818,PhysRevB.48.17006}. This justifies the numerical accuracy of our estimated $U$ and $J$ since calculated band-gap depends on the choice of $U$~\cite{pandaNiS,PhysRevB.89.155102}. A large exchange splitting between the two spin-channels of Fe-$3d$ is evident from the partial DOS (PDOS) in Fig.~\ref{VB_Structure_DOS_RPES}(d). One of the spin-channels of Fe-$3d$ is found to be completely occupied, while the other channel is completely empty, opening up a charge transfer gap between the O-$2p$ and Fe-$3d$ states.

\begin{figure*}[t]
\includegraphics[width=0.99\textwidth]{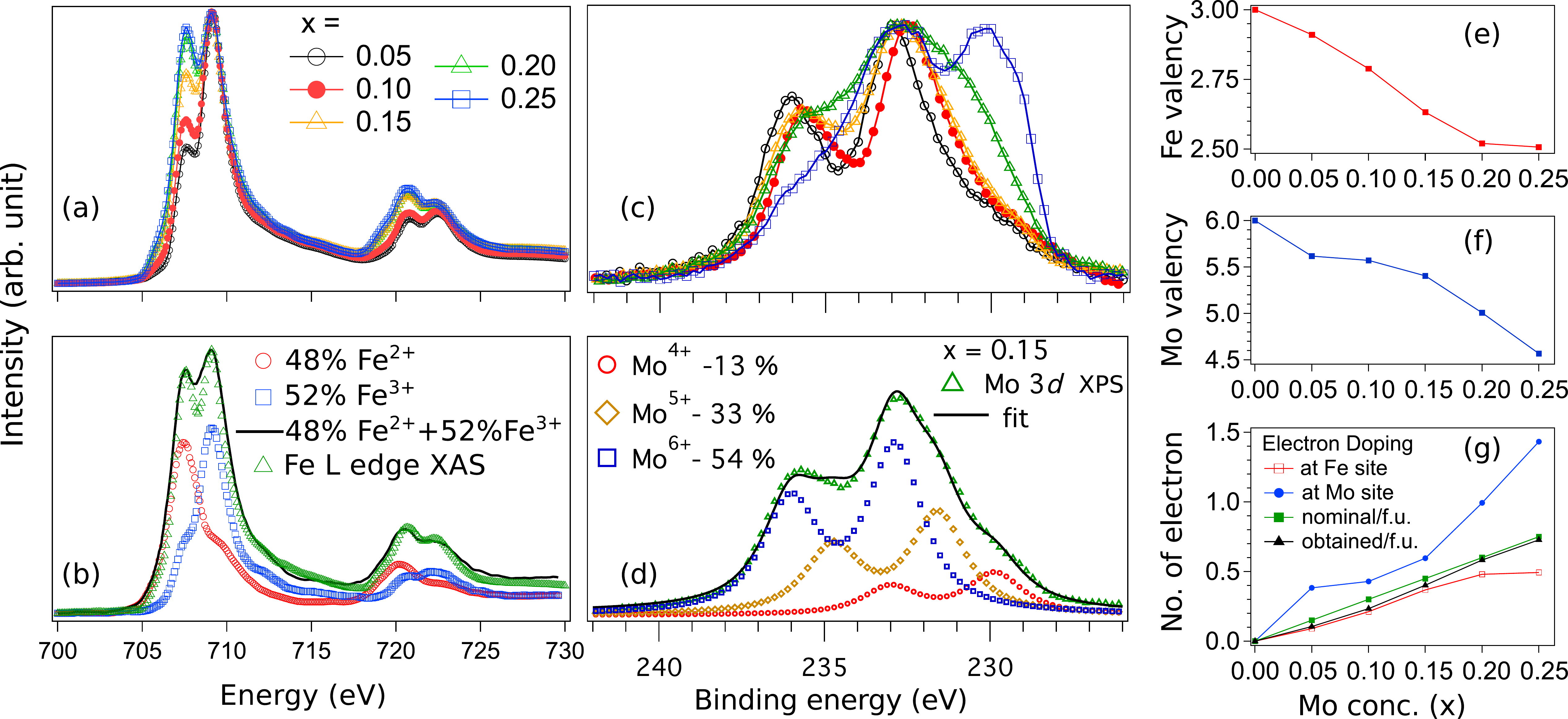}
\caption{{\bf Fe $L$ edge X-ray absorption and Mo $3d$ core level XPS spectra for valency determination.}(a) Fe $L$ edge XAS spectra for all the compositions. (b) Fitted data for $x$ = 0.20 sample by the weighted average of the reference spectra of Fe$^{2+}$ and Fe$^{3+}$. (c) Mo $3d$ core level XPS spectra. (d) Experimental (open triangle), and fitted (solid line) spectra for $x$ = 0.15 and the contributions from Mo$^{6+}$, Mo$^{5+}$ and Mo$^{4+}$. (e) Valency of Fe. (f) Valency of Mo. (g) Different contributions of doped electrons obtained at Fe band, Mo band, the total/f.u.\ and nominal/f.u.\ are shown.}
\label{Fe2pXAS_Mo3dXPS_Valency}
\end{figure*}

A major modification of the electronic structure is observed upon Mo doping as displayed in Fig.~\ref{VB_Structure_DOS_RPES}(d) and (e), for 12.5\% and 25\% dopant concentrations, respectively.  The total DOS corresponding to the 12.5\% Mo doping shows a strong reduction of the band gap to 0.23 eV. The occupied part of the Mo-$d$ states appears just below the Fermi level, which is exactly what we observe in experiment with 10\% and 15\% Mo doped samples (see Fig.\ \ref{Fig1}(b)), resulting in a small $d$-$d$ gap insulating ground state.
Upon increasing the Mo concentration to 25\%, the gap disappears and the system undergoes a doping driven IMT, lending strong support to our experimental finding. In the metallic phase, the Mo-$d$ states primarily contribute to the spectral weight at the Fermi level, while the occupied Fe-$3d$ states remain localised, located at several eV below the Fermi level, giving rise to the realization of a novel site-selective Mott insulating (SSMI) state.  We note that SSMI phase has recently drawn huge attention in rare-earth nickelates~\cite{SSMS_Nickelate} and also in Fe oxide~\cite{PhysRevX.8.031059} since it underlines a completely new mechanism, arising from the interplay of electronic correlation and lattice effects to describe IMT in strongly correlated electronic systems~\cite{SSMS_Nickelate}.

Further the PDOSs of two types of Fe show two different charge states as expected from the analysis of the crystal geometry. We find that the $3d$-states of the Fe1 are very similar to that of the pure LaFeO$_3$, implying a 3+ charge state for Fe1, while Fe2 exhibits distinctly different electronic structure. The majority $3d$-states of Fe2 are completely occupied, while the minority states are partially filled. This is in contrast to the Fe1 where the minority channel is completely empty, giving rise to a $2+$ (3$d^6$) S=2 high-spin state for Fe2. Thus the electronic structure confirms the CD state of Fe ions in 25\% Mo doped LaFeO$_3$, as was indicated from the observed breathing-distortion.

The theoretical finding is complemented by probing the orbital characters of various VB features using resonant PES (RPES) at the Fe $L_3$ and Mo $L_3$ absorption edges for the $x$ = 0.25 sample which are shown in Fig.\ \ref{VB_Structure_DOS_RPES}(f), and (g), respectively. The off and on resonance VB spectra are collected at 700 eV (2500 eV) and 709.9 eV (2522 eV) for Fe $L_3$ (Mo $L_3$) edges, respectively. No noticeable resonance effects are observed at $E_F$ for Fe, while a large enhancement of the spectral weight at $E_F$ in the spectra recorded at the Mo $L_3$ resonance photon energy is evident. This indicate that the Mo $d$ states are located around $E_F$, which is consistent with the calculated spectra of Fig.~\ref{VB_Structure_DOS_RPES}(e). Comparing the PDOS with the RPES, we could also identify different Fe features as follows: dominant Fe2 $e_g$ states are between 1-3 eV; hybridized Fe and O states are between 3-6.5 eV and Fe1+Fe2 $t_{2g}$ states are between 6.5-9 eV binding energy (compare Fig.\ \ref{VB_Structure_DOS_RPES}(e) with Fig.\ \ref{VB_Structure_DOS_RPES} (f) and (g)).

To verify the theoretically proposed CD ground state, Fe and Mo valency have been probed. We probe the doping dependence of the valency of the Fe ions by means of Fe $L$ edge X-ray XAS as displayed in Fig.~\ref{Fe2pXAS_Mo3dXPS_Valency}(a). Both the $L_3$ and $L_2$ edges exhibit a particular trend in the spectral features, while being more prominent at the $L_3$ edge. The $L_3$ edge is composed of two peaks. As evidenced from the reference spectra in Fig.\ \ref{Fe2pXAS_Mo3dXPS_Valency}(b)~\cite{vanAken1998}, the lower energy peak is predominantly Fe$^{2+}$ while the peak at higher photon energy is attributed to Fe$^{3+}$. A weighted sum of the reference spectra of Fe$^{2+}$ and Fe$^{3+}$ is used to model and fit the experimental data to extract the average valency of Fe for each composition. The model spectrum for the 20\% Mo doped sample (Fig.\ \ref{Fe2pXAS_Mo3dXPS_Valency}(b)) exhibits excellent agreement with all the experimental features although it is slightly broadened due to poorer spectral resolution of the reference spectra. We note here that our calculated XAS (see SI) by means of a combination of density functional theory and multiplet ligand-field theory nicely reproduces the observations shown in Figs.\ \ref{Fe2pXAS_Mo3dXPS_Valency} (a) and \ref{Fe2pXAS_Mo3dXPS_Valency} (b).

The valency of Mo is probed via $3d$ core level PES. Fig.\ \ref{Fe2pXAS_Mo3dXPS_Valency}(c) shows the background corrected spectra, normalized to match at maximum intensity, for all compositions. The spectra are fitted considering different valencies. A satisfactory description of the spectral features is achieved only when three Mo valencies, 6+ , 5+, and 4+, are taken into account. Fig.\ \ref{Fe2pXAS_Mo3dXPS_Valency}(d) shows the fitting and the corresponding contribution from different valency of Mo for the $x$ = 0.15 sample, while the same analysis for the other compositions and the details of the fittings are presented in the SI. 

The average valency of a Fe and Mo ions derived from the fits is plotted in Fig.\  \ref{Fe2pXAS_Mo3dXPS_Valency}(e) and (f), respectively. Three electrons are donated to the system when a Fe atom ($[Ar]3d^{6}4s^{2}$) is replaced by Mo ($[Kr]4d^{5}5s^{1}$). Fe is in a $3+$ valency ($[Ar]3d^{5}4s^{0}$) in the parent compound, while Mo is most stable in $6+$ valency ($[Kr]4d^{0}5s^{0}$) when all the outer electrons are donated.
The obtained numbers of doped electrons at the Fe and Mo sites relative to their most stable valency of $3+$ and $6+$, respectively, are plotted in Fig. \ref{Fe2pXAS_Mo3dXPS_Valency}(g). The total number of doped electrons/formula unit, i.e., the sum of doped electrons at Fe and Mo sites per formula unit, match the nominal values, (see Fig. \ref{Fe2pXAS_Mo3dXPS_Valency}(g)) demonstrating the accuracy of our employed techniques.

With increasing $x$, electrons are filled in both Fe and Mo bands (Fig.\ \ref{Fe2pXAS_Mo3dXPS_Valency}(g)). However, the electron doping rate at the Fe sites reduces after $x$ = 0.15 and saturates to $\sim$0.5/f.u. for $x\geq$0.20. As a result, a large increase of electron doping at the Mo sites is observed for the $x$ = 0.20 sample. Further inclusion of the Mo ions, i.e.\ as for the case of the $x$ = 0.25 sample, pushes all additional charge to the Mo sites. This is evident from both Fe-$2p$ XAS and Mo-$3d$ PES. The Fe-$2p$ XAS remain almost identical for the $x$ = 0.20 and 0.25 samples (Fig.\ \ref{Fe2pXAS_Mo3dXPS_Valency}(a)), while the Mo-$3d$ PES show a sudden increase of the spectral weight around 330 eV corresponding to a relative increase of the $4+$ state of Mo (Fig.\ \ref{Fe2pXAS_Mo3dXPS_Valency}(c)). These observations imply that the electron doping at the Fe sites reaches the percolation threshold when the valency becomes $2.5+$ beyond 20\% Mo doping. The $2.5+$ valency of Fe may imply equal sharing of one electron between two Fe-sites. However, as both our theoretical and experimental findings suggest, the Fe sector to be insulating, the electrons at the Fe-sites are localized, which implies an equal number of Fe$^{+3}$ and Fe$^{+2}$ ions in the system for the $x$ = 0.25 sample. 

\begin{figure}[t]
\includegraphics[width=0.99\columnwidth]{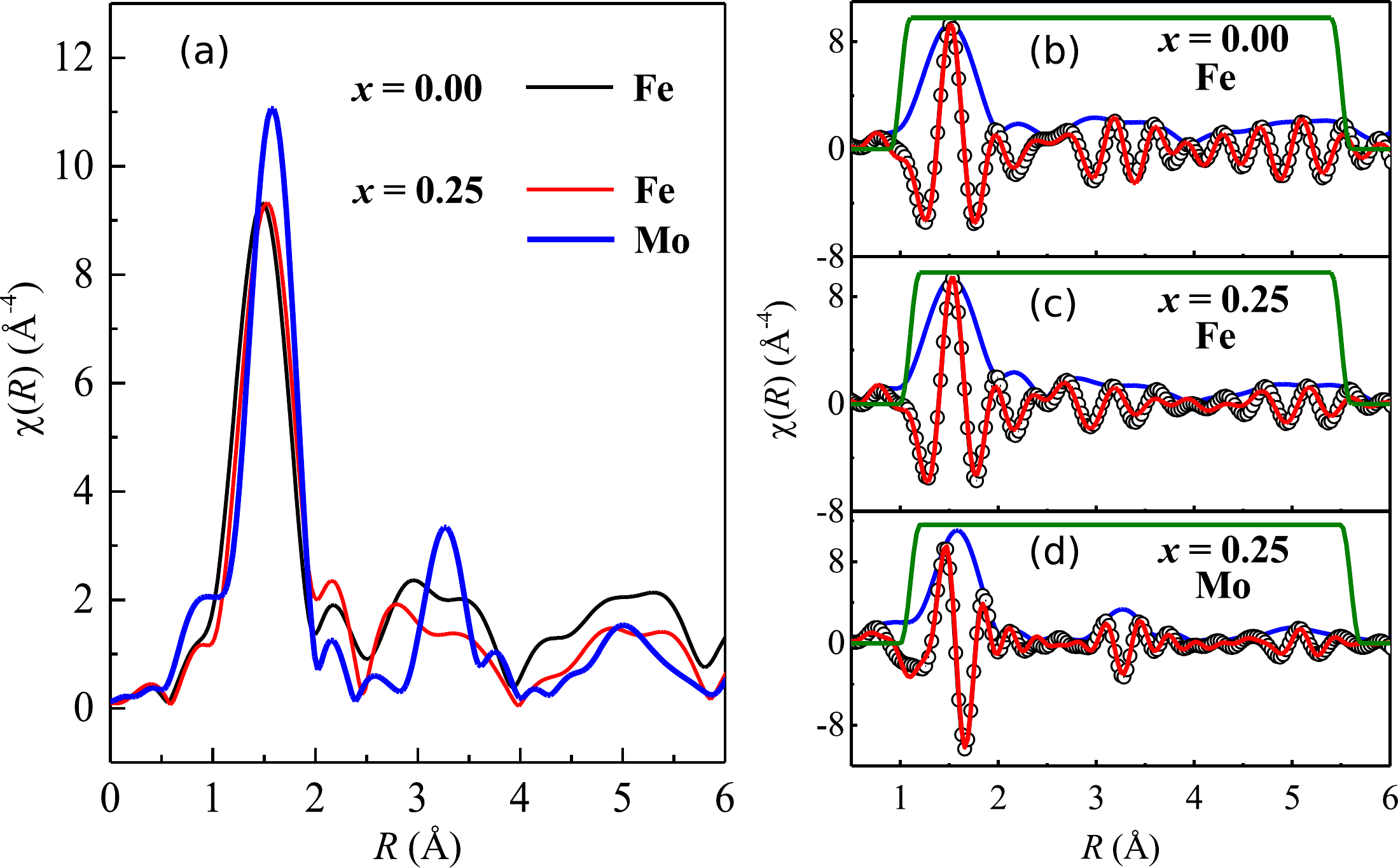}
\caption{{\bf X-ray absorption fine structure} (a) Magnitude of $\chi$($R$) for Fe-K EXAFS ($x$ = 0.0, black and x = 0.25, red) and Mo-K EXAFS ($x$ = 0.25, blue) collected at room temperature. Fits (red curves) to the real part of $\chi$($R$) (black open circles) for (b) Fe ($x$ = 0.0), (c) Fe ($x$ = 0.25) and (d) Mo ($x$ = 0.25). The corresponding magnitude of $\chi$($R$) for each data is presented in blue. The Hanning window function (olive) has been used over the range 1.1 $\AA$ to 5.5 $\AA$ for the fits.}
\label{Fig_EXAFS}
\end{figure}

Further the decisive test of the two types of Fe-O bond lengths in 25\% Mo doped sample has been done by X-ray absorption fine structure (XAFS) measurements at RT (see SI for details). The magnitude of Fourier transforms to the chi(R) data for Fe-K and Mo-K EXAFS and the fits to the corresponding real part of the chi(R) data are shown in Fig. \ref{Fig_EXAFS}. Details of the local parameters extracted from EXAFS fitting are presented in Table II in SI. The two Fe-O distances, which in pure LaFeO$_3$ differs by only 0.024 \AA, has been increased to 0.044 \AA in 25\% Mo doping, which is in excellent agreement with the theoretical value of 0.05 Å. Therefore, two types of Fe are present even at RT, where the conduction is established by the partially filled Mo $t_{2g}$ bands, providing further evidence in favor of a unique site-selective Mott phase with the CD AFM state.

%\section*{Discussion}
Using a combination of experimental and theoretical investigations, we unambiguously conclude that Mo-doped LaFeO$_3$ possesses a unique state at higher doping levels, where antiferromagnetism and metallic bands coexist at RT with an associated charge disproportionation at the Fe-sites. The Fe sector remains localized having two types of Fe in the unit cell: Fe$^{3+}$ and Fe$^{2+}$, while the metallic character is attributed to the partially filled Mo $t_{2g}$ bands. The lattice is highly sensitive to surface defects.
Mo doped LaFeO$_3$ at $x$ = 0.25 is therefore a particularly exotic example of a site-selective Mott insulator, where itinerant and localized orbitals reside on different atomic sites and the transition to the localized state on the Fe sites is driven by the CD. On general grounds, in such a system, exotic metallic behavior (including non-Fermi liquid properties)~\cite{PhysRevLett.95.206401} could be expected if the pure bulk intrinsic resistivity was accessible to experimental measurements.

\section*{Methods}
Experiment: Six samples of LFMO-x ($x$ = 0.00, 0.05, 0.10, 0.15, 0.20, 0.25) are synthesized by a solid state synthesis route (see supplementary information (SI)). The phase purity of the samples was checked by X-ray diffraction (XRD) using Cu $K_{\alpha , \lambda} = 1.5406$ {\AA}. Magnetic measurements were carried out using a Quantum Design SQUID magnetometer. Resistivity measurements were performed using a four probe setup. Scanning tunnelling spectroscopy (STS) measurements were done at room temperature in an UHV system, using a standard ac modulation technique with voltage modulation of 10mV, 2731 Hz sine wave. X-ray photoelectron spectroscopy (XPS) and X-ray absorption spectroscopy (XAS) measurements were performed at the D1011 and I1011-beamlines of the Swedish synchrotron facility MAX-lab, Lund. To eliminate surface contamination, samples surfaces were cleaned in vacuum before recording  XAS and XPS spectra. XAS spectra were measured by recording the total electron yield. Room temperature X-ray absorption fine structure (XAFS) measurements were performed at the Fe-K edge and Mo-K edge at the Sector 10 bending magnet beamline at the Advanced Photon Source, Argonne National Laboratory, Chicago.
\par
Electronic structure calculations: On the theoretical side, we first calculate the non-spin polarized DFT-GGA~\cite{GGA_PBE} (generalized gradient approximation) electronic structure by means of the Wien2k code~\cite{wien2k}, based on the full potential linearized augmented plane wave (FP-LAPW) method~\cite{FPLAPW_Wien2k}. The charge self-consistent converged GGA electronic structure is also the starting point for the calulation of the  effective interaction parameters (Hubbard $U$ and Hund's $J$) within the constrained random phase approximation (cRPA)~\cite{cRPA_orig,cRPA_wien2kImple} method. The rest of the analyses is carried out within the GGA+$U$ approach using the estimated $U$ and $J$ for Fe-$d$ states which are respectively 4.4 eV and 0.7 eV. We have considered three structures: 25\% Mo doped, 12.5\% Mo doped, and pure LaFeO$_3$ to understand our experimental results. The optimization of all the atomic coordinates are performed using the plane-wave based method as implemented in the Vienna ab initio simulation package (VASP)~\cite{vasp2}. Further these optimized structures are used to calculate the electronic structure using FP-LAPW method~\cite{FPLAPW_Wien2k}. The other technical details  are provided in SI.

%\bibliography{main}


\section*{Acknowledgments}
A.D. acknowledges SERB, India, for National Post-Doctoral Fellowship. O.E. acknowledges the support from Swedish Research Council (VR), eSSENCE, STandUPP, the Knut and Alice Wallenberg (KAW) Foundation and the foundation for strategic research (SSF). MVK acknowledges the VR starting Grant (No. 2016-03278) from the Swedish Research Council. DDS thanks Jamsetji Tata Trust for support. MRCAT operations are supported by the Department of Energy and the MRCAT member institutions. This work was supported  by  a Consolidator  Grant  of  the  European  Research  Council  (Project  No.   617196)  and  supercomputing  time  at
IDRIS/GENCI Orsay (Project No. t2018091393). SKP and SB thank the computer team of Centre de Physique Th\'eorique for support.

\section*{Author contributions}
SJ conceived the project, planned the experiments, performed magnetization, resistivity, XPS, XAS experiments and analyzed the data. The DFT+U calculations and its analysis were done by SKP, SB and OE. DP performed XPS, RPES experiments and their analysis. SJ and BP synthesized and characterized the samples. SM and CUS have performed the XAFS measurements and analysis. AD and DDS planned and executed STS measurements and its analysis. PAK, DH, PS, KG have performed the magnetic measurements. JS, PT, YV, OE have done the XAS spectra calculations and its analysis. MVK have done resistivity measurements. OK has supported and supervised SJ for performing the experiments at various facilities. The manuscript was largely written by SJ and SKP, DDS did the major revision, suggestions and comments of all the co-authors are incorporated. SJ, DDS, SB, SKP have significant contribution in interpreting the results. 

%\section*{Additional information}
%\textbf{Supplementary information} accompanies this paper at https://www.nature.com/nphys/ \\
%\textbf{Competing financial and/or non-financial interests:} The authors declare no competing financial and or non-financial interests.
\end{document}